\documentclass[twocolumn]{article}
\usepackage{graphicx,amsmath,amssymb,longtable,url}
\usepackage[latin9]{inputenc}

  \renewenvironment{thebibliography}[1]{%
    \begin{oldthebibliography}{#1}%
      \setlength{\parskip}{0ex}%
      \setlength{\itemsep}{0ex}%
  }%
  {%
    \end{oldthebibliography}%

  }

\newcommand{\be}{\begin{equation}}
\newcommand{\ee}{\end{equation}}
\newcommand{\bea}{\begin{eqnarray}}
\newcommand{\eea}{\end{eqnarray}}

\newcommand{\ca}{a}

\renewcommand{\a}{{\bf a}}
\renewcommand{\b}{{\bf b}}

\newcommand{\bth}{{\bf \Theta}}

\newcommand{\A}{\pi}

\newcommand{\opone}{\leavevmode\hbox{\small1\kern-3.3pt\normalsize1}}
\newcommand{\para}[1]{\vspace{1ex} \noindent {\bf #1}}

\begin{document}

\title{Cross-species analysis of biological networks by Bayesian alignment}

\author{Johannes Berg and Michael L\"assig \\
Institut f\"ur Theoretische Physik,
Universit\"at zu K\"oln \\ 
Z\"ulpicherstr. 77, 50937 K\"oln, Germany\\
}
\maketitle

\vspace{10 pt}

{ \bf \noindent
Complex interactions between genes or proteins contribute a substantial part
to phenotypic evolution. Here we develop an evolutionarily grounded
method for the cross-species analysis of interaction networks
by {\em alignment}, which maps bona fide functional relationships
between genes in different organisms. Network alignment is based on a
scoring function measuring mutual similarities between networks taking into account 
their interaction patterns as well as sequence similarities between
their nodes.  High-scoring alignments and optimal alignment
parameters are inferred by a systematic Bayesian analysis. We apply
this method to analyze the evolution of co-expression networks between
human and mouse. We find evidence for significant conservation of gene
expression clusters and  give network-based predictions of gene function. 
We discuss examples where cross-species functional relationships between genes 
do not concur with sequence similarity.  
}

\vspace{1ex}


Besides a wealth of genomic sequence information, molecular biology is
accumulating more and more data probing the interactions between genes
or proteins. Examples are regulatory interactions, where the
expression level of one gene influences the expression of another
gene, or interactions between proteins, where pairs of proteins bind
to form dimers or multimers. Interactions between
genes or their products are crucial for our understanding of
biological functions. With the advent of experimental high-throughput
methods, large-scale datasets of different organisms are becoming
available, which can be analyzed by systematic cross-species
comparison.

This paper is devoted to developing an evolutionary rationale for
biological network analysis. Since the interactions between genes are
encoded in their genomic sequences, this may seem a rather
straightforward generalization of established concepts in sequence
analysis: evolution acts as a divergent process on the
constituents of the network, which gradually reduces cross-species
correlations of the network structure. Detecting these correlations
requires an alignment procedure which can map functional units as
network structures conserved by evolution as well as estimate the
degree of divergence between species.

However, interaction networks evolve in a more {\em heterogeneous} and a more 
{\em correlated} way than sequences, which makes their cross-species
comparison a considerably more challenging task. The interactions between
proteins, for example, depend on the properties of a specific functional
binding domain, which may evolve in a different way than the remainder of the
protein sequence, with correlations to its binding domain in a different 
protein.   Regulatory
interactions can change by the evolution of regulatory DNA, which is expected
to be different from that of coding DNA. Moreover, many sequence changes in a
gene may be irrelevant for its interactions measured in a network.  This leads
us to treat the evolution of the interactions within a network, i.e., its {\em
link dynamics}, and the overall sequence evolution of its constituents, the
{\em node dynamics}, as two independent modes of evolution. We describe these
modes by simplified stochastic models and infer their relative contribution to
network evolution by cross-species comparison. This dynamics is also quite
heterogeneous across the network, and we use the models of link and node
dynamics to quantify the evolutionary conservation of putatively functional
network modules.

Our evolutionary analysis is based on the {\em alignment} of networks, i.e., 
a mapping between their nodes, which also induces a mapping of their links. 
In the Theory part of this 
paper, we develop a statistical theory of alignment for biological networks.
We introduce a scoring designed to detect local functional
correlations, which uses both the similarities of mapped link pairs and of  
node pairs. This scoring derives from the underlying link and node
dynamics. 

Various alignment and scoring procedures for biological networks have
been discussed in recent articles. One type of methods restricts the
alignment to mutually homologous nodes, i.e, gene pairs with
significant sequence similarity in different species. In this way,
clusters of conserved interactions have been found in gene
co-expression
networks~\cite{StuartSegalKollerKim:2003,BergmannIhmelsBarkai:2004}
and in protein interaction
networks~\cite{SharanSuthrametal:2005,SharanIdekeretal:2005}.  A
complementary approach is to align networks only by their link
overlap, independently of node homology. Network
motifs~\cite{Shen-Orr.etal:2002,Milo.etal:2002} defined by families of
mutually similar subgraphs in a larger network have been identified in
this way~\cite{BergLaessig:2004} as well as the similarities between
regulatory networks of different phages~\cite{Trusinaetal:2005}. These
methods have been combined with their relative weights fixed ad hoc in
ref.~\cite{HeymansSingh:2003}.  A third method called
Pathblast~\cite{KelleySharan:2003,KelleyYuan:2004} evaluates the link
similarity between networks along paths of connected nodes, using
sequence alignment algorithms. It has been applied to cross-species
comparisons of protein interaction
networks~\cite{KelleySharan:2003}. Similarly, the flux along the
shortest paths in regulatory networks has been compared across
species~\cite{Trusinaetal:2005}.  Metabolic networks with few cycles
have been analyzed by subtree comparison~\cite{Pinteretal:2005}.

From an evolutionary point of view, these methods are heuristics containing
different assumptions on the underlying link and node dynamics. Homology-based
alignments are appropriate if the sequence divergence between the species
compared is sufficiently small so that all pairs of functionally related 
nodes can be mapped by sequence homology. However, genes with entirely 
unrelated sequence may take on a similar function in different organisms, 
and hence have a similar position in the two networks. (Such so-called 
non-orthologous gene displacements are well-known in metabolic 
networks~\cite{KooninMushegianBork:1996,GalperinKoonin:1998,Morett_etal:2004}.)
On the other hand, alignments by link
similarity alone altogether ignore the evolutionary information of the 
node sequences. Path-based alignment algorithms are well suited to networks
with predominantly linear biological pathways 
such as signal-transduction chains. In other
situations, however, it may be difficult to link the scoring parameters to
evolutionary rates of link and node changes. 

The alignment method  presented in this paper is grounded on statistical models
for the evolution of links and nodes. Alignments are constructed from link and
node similarity treated on an equal footing, the relative weight of these score
contributions is determined systematically by a Bayesian parameter inference. 
Nodes without significant sequence similarity are aligned
if  their link patterns are sufficiently similar. Conversely, nodes are not 
aligned despite their sequence similarity, if their links, and hence
their putative functional role, show a strong divergence between the two 
networks. Our method is rather general and can be applied both to networks
with binary link strengths (as in the current large-throughput data for protein
interactions) and to networks with  continuous link strength (such as the
co-expression data used in this study).

As an algorithmic problem, network alignment is clearly more challenging
than sequence alignment, which can be solved by dynamic
programming~\cite{NeedlemanWunsch:1970,SmithWaterman:1981}. 
Already simpler problems such as matching two graphs by determining
the largest common subgraph are $NP$-hard~\cite{Papadimitriou}, which
implies there is probably no polynomial-time algorithm.  We have
developed an efficient heuristic, by which network alignment is mapped
onto to a generalized quadratic assignment problem, which in turn can
be solved by iteration of a linear problem~\cite{Tsafriretal:2005}.

\enlargethispage{\baselineskip}
In the second part of the paper, we present a cross-species comparison of
co-expression networks of {\it H. sapiens} and {\it M. musculus} as an
example application of our method. In this type of networks, the link
between a pair of genes is given by the correlation coefficient of their
expression profiles measured on an mRNA microarray chip.
We show that
correlation networks are well-suited for cross-species comparison: they are
robust datasets even if  individual expression levels cannot be compared with
each other since  the experimental conditions differ between species. The
evolution of  these networks results from the evolution of  regulatory
interactions between genes and from loss and gain of genes. High-scoring
alignments between expression networks in human and mouse provide a
quantitative measure of divergence between the two species. We find conserved
network structures, related to clusters of co-expressed genes; similar
findings are reported in refs.~\cite{StuartSegalKollerKim:2003}
and~\cite{SharanIdekeretal:2005}. However, the alignment 
found here differs from mere sequence homology. This
leads to network-based predictions  of gene functions, including functional
innovations such as non-orthologous gene displacements. 

\subsection*{Theory}

\para{Graphs and graph alignments.}
A {\em graph} $A$ is a set of {\em nodes} with {\em links} between pairs of nodes.
The graphs considered here are labeled by gene name, which is 
denoted by the node index $i= 1, \ldots, N_A$, and are thus uniquely 
represented by the {\em adjacency matrix} $\a = (a_{ii'})$.  A graph is called
{\em binary} if links are   either absent ($\ca_{ii'}=0$) or present
($\ca_{ii'} = 1$) and {\em weighted}  if the link strengths 
$\ca_{ii'}$ take continuous values.   The
special case of a symmetric adjacency matrix is used to describe {\em
undirected} graphs. 
For example, current high-throughput datasets of protein
interactions, which do not specify the interaction strength, produce binary 
undirected graphs. Gene expression networks, whose links denote the mutual
correlation coefficient between expression patterns of two genes, 
are weighted graphs with $-1 \leq \ca_{ii'} \leq 1$. 

\begin{figure}[tb!]
\includegraphics*[width=  \linewidth]{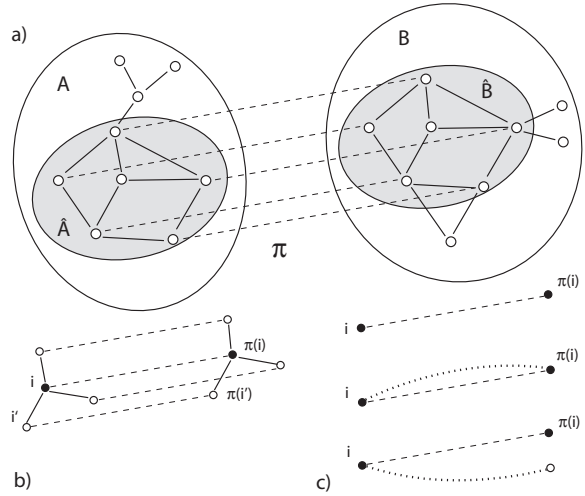}
\caption{\label{fig_stat} \small
{\bf Network alignments measure link and node similarity.} 
(a)~A local alignment $\pi$ between two networks $A,B$ is a one-to-one mapping 
(indicated by dashed lines) between 
nodes of the subsets $\hat A$, $\hat B$. 
(b)~The local link score 
$S^\ell_{i, \pi(i)}$ evaluates all pairwise
similarities between links $a_{ii'}$ and $b_{\pi(i) \pi(i')}$ (solid lines) for a given pair 
of aligned nodes. 
(c)~The local node similarity score $S^n_{i, \pi(i)}$ evaluates the overlap
of the alignment with the node similarity $\theta_{i, \pi(i)}$ (dotted line). 
Top to bottom: Aligned node pairs 
(i) without similarity 
to any other node,
(ii) with mutual node similarity,
(iii) with (at least one) node similarity mismatch. 
}
\end{figure}

A {\em local alignment} between two graphs $A$ and $B$ is defined as a
mapping $\pi$ between two subgraphs $\hat A \subset A$ and $\hat B =
\pi(\hat A) \subset B$ as shown in Fig.~\ref{fig_stat}(a).  The
alignments of networks discussed in this paper are designed to display
cross-species functional relationships between aligned node pairs. Due
to gain or loss of genes in either species, not every gene in one
network has a functional equivalent in the other, and the alignment
algorithm has to determine the aligned subnetworks $\hat A$ and $\hat
B$ with significant correlations. For the sake of algorithmic
simplicity, we will discuss here only one-to-one mappings $\pi$, which
is appropriate for most gene pairs but neglects multi-valued
functional relationships induced for some genes by gene duplications.

\para{Link dynamics and link score.}
An important statistical characteristic of a network is the link
distribution $p^\ell (\ca)$, giving the probability that the link 
between a randomly chosen pair of nodes takes on the value $a$. 
The evolution of the link distribution can be modeled
by a simple stochastic process, from which our link similarity scoring 
of an alignment is derived. 
In a binary network, the simplest form of link dynamics is a Markov
process, which is fully determined by the rates of formation and loss of  
single links. Generalizing this dynamic to continuous links leads to a 
diffusion equation of the form
\begin{equation}
\partial_t \, p^\ell (\ca) = [\partial_\ca^2 g(a) - \partial_\ca f(\ca)] 
  \, p^\ell (\ca).
\label{diff}
\end{equation}
The two terms on the r.h.s.~describe the stochastic turnover and the average
relaxation of links with coefficient functions $g(a)$ and $f(a)$, respectively. 
For mutual expression correlations between two
genes in a microarray, this form can be derived from a stochastic model for
loss and gain of regulatory interactions, each of which affects a random
subset of the experiments, resp.~cell types. 
The cross-species correlations in pairs of 
evolutionarily related links $a,b$ 
are contained in the joint distribution $q^\ell(a,b)$,
which we write in the form
\be
q^\ell (a,b) = p^\ell_A (a) \, p^\ell_B (b) \exp[s^\ell (a,b)],
\label{Sl}
\ee
defining the log-likelihood {\em link similarity score} $s^\ell (a,b)$. 
For binary links, this has a bilinear form
\be 
\label{Slc}
s^\ell (a,b) = \lambda_\ell \, a b  + \sigma^\ell_A a + \sigma^\ell_B b + {\rm const.},  
\ee
with the link match reward $\lambda_\ell$ depending on the evolutionary 
distance between the species.
The additive constant is given by the
normalization of the probability distributions in (\ref{Sl}). 
For continuous links, we write the joint distribution as 
$q^\ell (a,b) = G(b|a)\, p_A^\ell (a)$, where $G(b|a)$ is the conditional distribution of 
link strengths $b$ evolved from an initial strength $a$ over the evolutionary 
distance between the two species compared. For short evolutionary distances and 
for a link evolution of the form (\ref{diff}), this distribution is well 
approximated by a Gaussian, 
$G(b|a) \sim \exp[ - \lambda_\ell \, g(\frac{a+b}{2})\, (a - b)^2]$. For large
evolutionary distances, it can be shown that the score 
$s^\ell (a,b) = G(b|a) / p^\ell_B (b)$ has again the asymptotic form (\ref{Slc}).  
Given datasets of two networks $A$ and $B$, the distribution $p_A^\ell$, $p_B^\ell$ can be
estimated from the frequency of link strengths $a_{ii'}$ resp. $b_{jj'}$ in
one species, and $q^\ell$ from the frequency of link pairs $(a_{ii'}, b_{jj'})$
involving orthologous gene pairs $(i,j)$ and $(i',j')$. Hence, the score
function $s^\ell(a,b)$ defined by eq.~(\ref{Sl}) can be inferred without specific
assumptions on the underlying link dynamics. For the example discussed below,
this empirical link score turns out to be in 
remarkable agreement with the form (\ref{Slc}) predicted by the link diffusion 
model. 

This scoring of individual link pairs is readily generalized to pairs of 
networks $(A,B)$ with a given local alignment $\pi$. Assuming that aligned
link pairs $(a_{ii'}, b_{\pi(i) \pi(i')})$ follow the distribution $q^\ell(a,b)$ 
independently from each other and unaligned links $a_{ii'}$ and $b_{jj'}$
follow the distributions $p_A^\ell(a)$ and $p_B^\ell(b)$, respectively, 
we obtain the distribution of graph pairs for 
given $\pi$, 
\be
\label{Ql}
Q^\ell (\a,\b|\pi) = P_A^\ell (\a) P_B^\ell (\b) \exp[S^\ell (\a,\b, \pi)],
\ee
where $P_A^\ell (\a) = \prod_{i,i' \in A} p_A^\ell (a_{ii'})$, 
$P_B^\ell (\b)$ has a similar product form, and the network link
score $S^\ell (\a,\b,\pi)$ is a sum of local contributions 
$S^\ell_{i,\pi(i)}$ of aligned node pairs,
\be
\label{Sl_total}
S^\ell (\a,\b,\pi) = \sum_{i \in \hat A} S^\ell_{i,\pi(i)}
  = \sum_{i,i' \in \hat A} s^\ell (a_{ii'}, b_{\pi(i) \pi(i')}),
\ee
as shown in Fig.~\ref{fig_stat}(b). 
For co-expression networks, there are correlations
between links within one network. These occur since the number of independent measurements,
$d$, is smaller than the number of genes $N$, and are taken into account by the overall
scale of the link score (i.e, $\lambda_\ell \sim \sigma^\ell \sim d/N$). 

The relative evolutionary conservation
of a given pair $(a,b)$ of aligned links within the network is measured
by its excess link score
\be
\label{excess_link_score}
\Delta s^\ell(a,b) \equiv s^\ell(a,b) - 
( \langle s^\ell(a,b') \rangle_{b'} + \langle s^\ell(a',b) \rangle_{a'} )/2,
\ee
i.e., the difference of its link score and the average over all aligned 
link pairs with either strength $a$ fixed, 
$\langle s^\ell(a,b') \rangle_{b'} \equiv \int d\,b' G(b'|a) s^\ell(a,b')$,
or strength $b$ fixed.  The relative conservation of link patterns 
between a pair of aligned nodes $i,\A(i)$ is then given by 
 the 
local excess link score 
\be
\label{local_excess_link_score}
\Delta S^\ell_{i,\pi(i)} = \sum_{i' \in \hat{A}} \Delta s^\ell(a_{ii'},b_{\pi(i)\pi(i')}) \ .  
\ee
These measures will become important for the identification of network 
clusters and their evolutionary conservation. 
 
\para{Node dynamics and node score.}
The pairwise similarity between genes in networks $A$ and $B$ is given by a
matrix $\bth$, whose entries $\theta_{i j}$ quantify, for
example, the overall sequence similarity between the gene sequences $i \in A$  and
$j \in B$ or a biochemical similarity between the corresponding proteins.  The
sequence similarity between functionally related genes decays over time due to 
local mutations, but is also affected by large-scale genomic events such  as
gene duplications, gene loss, or recruitment of new genes into a functional
context. Due to these processes, both networks contain a fraction of nodes with little or no
significant sequence similarity to any node in the other network, which  should
nevertheless be included in the alignment if their local link score suggests
significant functional cross-species relationships.  Moreover, {\em functional
swaps} between genes induce functional correlations between genes that are
unrelated by sequence and, at the same time, reduce correlations between other 
genes despite their sequence similarity. A prominent example is non-orthologous gene 
displacements~\cite{KooninMushegianBork:1996,GalperinKoonin:1998,Morett_etal:2004}.
It is these processes that cause the
network alignment to deviate for some nodes from a map based only on  sequence
homology. Functional swaps can be regarded as part of the link evolution, which
in co-expression networks leads to {\em coherent} link changes at the affected
nodes. However, these swaps are likely to involve selection and are not
captured by the independent link dynamics discussed above. Hence, we include
them here as a separate type of process with its own evolutionary rate.

The resulting statistics of node similarity can be described by  the
distribution of pairwise similarity coefficients between unaligned nodes,
$p_0^n (\theta)$, between pairs of aligned nodes, $q^n_1
(\theta)$, and between one aligned node and nodes other than its alignment
partner, $q^n_2 (\theta)$. 
Note that $p_0^n (\theta)$ does not simply describe uncorrelated
sequences: significant sequence similarity may exist between genes
that are not aligned due to their link mismatch, since  functional changes can
lead to a rapid divergence of links, for example, in the  formation of a
pseudogene.  

The  log-likelihood {\em node similarity scores}
$s^n_1 (\theta)$ and $s^n_2 (\theta)$, which are defined by 
\be 
\label{node_ensembles}
q^n_1(\theta) = p_0^n (\theta) \exp[s^n_1 (\theta)], \;\;\;
q^n_2(\theta) = p_0^n (\theta) \exp[s^n_2 (\theta)],
\ee
quantify the dependence of the alignment on node similarity. 
Assuming that the coefficients $\theta_{ij}$ are drawn independently
from these distributions, we obtain the distribution of node similarity 
for a pair of networks $A$ and $B$ with a given alignment $\pi$,
\be
\label{Qn}
Q^n (\bth| \pi) = P^n_0 (\bth) \exp[S^n (\bth, \pi)]
\ee
where $P^n_0 (\bth) = \prod_{i,j} p_0^n (\theta_{ij})$ and
the network node score $S^n (\bth, \pi)$ is again a sum of
local contributions
$s_1^n (\theta_{ij})$ and $s_2^n(\theta_{ij})$.
In this paper, we use a simple binary approximation of node 
similarity: two genes are counted as orthologous ($\theta_{ij} =1$)
if they appear as putative orthologs in the Ensembl-database~\cite{Ensembl:2005}, 
and otherwise not ($\theta_{ij} = 0$). Each node may have several such putative orthologs. 
The three distributions
in (\ref{node_ensembles}) are then fully determined by three model parameters, 
$p^n_0(\theta) \sim \exp[\zeta_0 \theta]$, 
$q^n_1(\theta) \sim \exp[(\zeta_0+\lambda_n) \theta]$, and 
$q^n_2(\theta) \sim \exp[(\zeta_0+\lambda_n') \theta]$, 
which in turn depend
on the rates of the node dynamics and on the evolutionary distance between 
the species. 
A short calculation shows that the node score~(\ref{Qn}) takes the form
\be 
\label{Sn_total}
S^n (\bth, \pi) = 
  \sum_{i \in \hat A} \big (
     S^n_{i, \pi(i)} +  \mu  \big ).
\ee
Here the local {\em node similarity score} 
\be
S^n_{i,\pi(i)} =\left\{ 
\begin{array}{ll}
0 & \mbox{if $\sum_{j\in B} \theta_{ij} = \sum_{i' \in A} \theta_{i' \pi(i)} = 0$,}
\\
\lambda_n & \mbox{if $\theta_{i \pi(i)} = 1$,} 
\\
\lambda_n' & \mbox{otherwise} 
\end{array}
\right.
\ee 
measures the overlap of alignment $\pi$ and homology 
map $\bth$ as shown in Fig.~\ref{fig_stat}(c), and  
the ``chemical potential'' $\mu (\lambda_n,\lambda_n', \zeta_0)$ implicitly 
determines the overall number of nodes in the alignment (for details, see the 
{\em Supporting Text}).
For large $\mu$, the highest scores occur
in {\em global alignments} between the networks $A$ and $B$, which involve 
all nodes of the smaller network. This is appropriate if the evolution 
of links and nodes maintains for all nodes some functional relationship within
the network. In the case of this study, link and node dynamics 
destroy significant correlations for some nodes. We obtain
local alignments with chemical potential $\mu < 0$, 
which exclude some nodes of both networks.  
              
\para{Hidden Markov model and Bayesian analysis.}
We can now combine the distributions $Q^\ell$ and $Q^n$ into a 
probabilistic model for link and node similarity, which produces the 
observable data, i.e., pairs of networks with adjacency matrices $\a$, $\b$ and
node similarity matrix $\bth$, 
for a given alignment $\pi$ and for given model parameters 
$m = (s^\ell, \lambda_n, \lambda_n',\zeta_0)$ in eqs.~(\ref{Sl_total}) and (\ref{Sn_total}). The combined model
is given by the probability distribution 
\begin{eqnarray}
\label{Q-model}
\lefteqn{Q(\a,\b,\bth | \pi, m) 
    =  Q^\ell (\a,\b,\bth | \pi, m)  
         Q^n (\bth | \pi, m)}\ \ \  \\
   && = \exp[S(\a,\b,\bth,\pi,m)] P_A^\ell (\a) P_B^\ell (\b) P_0^n (\bth,\zeta_0) \nonumber
\end{eqnarray}
with the alignment score function 
\be
S(\a,\b,\bth,\pi,m) = S^\ell (\a,\b,\pi,m) + S^n (\bth,\pi,m).
\label{Stot}
\ee
Eqs.~(\ref{Q-model}) and~(\ref{Stot}) are at the heart of our scoring procedure:
they provide a probabilistic rationale for the cross-species analysis of network 
data by link and node similarity. The model parameters $m$, which determine
the relative weight of link and node score, and the alignment $\pi$ are 
``hidden'' variables, which 
can be inferred by a standard Bayesian 
analysis. We write their posterior probability, i.e., the conditional 
probability of the hidden variables for given data $\a,\b,\bth$,
in the form
\be
Q(\pi,m | \a,\b,\bth) = \frac{Q(\a,\b,\bth | \pi,m) P (\pi, m)}{ 
   \sum_{\pi,m} Q(\a,\b,\bth | \pi,m)P (\pi, m) } 
\label{Qnet}
\ee
and assume the prior probability $P (\pi,m)$ to be flat. Dropping the terms
independent of $\pi$ and $m$, we obtain the optimal local alignment $\pi^*$ by 
maximizing the posterior probability 
$Q(\pi | \a,\b,\bth) \sim \sum_m Q(\a,\b,\bth | \pi,m)$ and 
similarly the optimal scoring parameters $m^*$ by maximizing  
$Q(m | \a,\b,\bth) \sim \sum_\pi Q(\a,\b,\bth | \pi,m)$. In a Viterbi
approximation, $\pi^*$ and $m^*$ can be inferred jointly by maximizing 
$ Q(\a,\b,\bth | \pi,m)$. This amounts to determining the optimal null model
parameter $\zeta_0$ and maximizing the combined score
$S(\a,\b,\bth,\pi,m)$. Details are given in the 
{\it Supporting Text}. 

\begin{figure}[tb!]
\includegraphics*[width=.7 \linewidth]{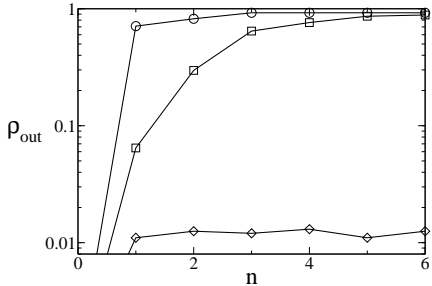}
\caption{\label{fig_proof_of_principle} \small
{\bf Performance characteristic of the alignment algorithm.} 
The fraction $\rho_{\rm out} $ of correctly aligned nodes is plotted 
against the number of iterative steps $n$ for fractions 
$\rho_{\rm in}=0.01$ (diamonds), $0.02$ (squares), $0.5$ (circles)  
of the node similarity given as input. 
Typically the algorithm converges after about $5$ iterations. 
There is a switch from low to 
high alignment quality ($\rho_{\rm out} > 0.9$) at a threshold 
value $\rho_{\rm c} \approx 0.02$.
}
\end{figure}

\para{Alignment algorithm.}
Our algorithm for maximizing the score is based on a mapping to a
generalized quadratic assignment problem, which is solved by an
iterative heuristic similar to~\cite{Tsafriretal:2005} with 
running times of order $N^3$~\cite{JonkerVolgenant:1987} (for details,
see {\em Supporting Text}). To quantify the performance of the
algorithm for co-expression networks, we have used a human microarray
dataset~\cite{Su.etal:2004}, consisting of expression measurements of
different tissues. We randomly partitioned the experiments into two
equally large subsets, and thus obtained two ``mirror copies'' of the
expression correlation network in one species. The nodes in the two
networks are identical and their links differ only by experimental
noise. The correct alignment of these two copies is trivial, $\pi (i)
= i$. A fraction $\rho_{\rm in}$ of correctly aligned nodes with
randomly chosen indices $i$ is given as input for the algorithm by
specifying the corresponding node similarity coefficients $\theta_{ij}
= \delta_{ij}$, the remaining node information is ignored
($\theta_{ij} = 0$). We then record the fraction of correctly aligned
nodes $\rho_{\rm out} (n)$ of the algorithm as a function of the
number of iterations $n$, see Fig.~\ref{fig_proof_of_principle}.  This
performance characteristic shows a switch from low to high alignment
quality at a threshold value $\rho_c \approx 0.02$. In the low-quality
regime ($\rho_{\rm in} < \rho_c$), the alignment contains for all $n$
only the nodes given as input.  In the high-quality regime ($\rho_{\rm
  in} > \rho_c$), the iterations continuously improve the fraction of
correctly aligned nodes, saturating at an accuracy $\rho_{\rm out} >
0.9$ for large $n$. Of course, the threshold will be higher and the
saturation accuracy lower for cross-species comparisons, where the
networks differ by evolutionary changes and by larger experimental
variation. Similarly, the threshold rises if the network is randomly
diluted (to $\rho_c \approx 0.2$ when $95\%$ of all links have been set to
zero).



\subsection*{Results}

\begin{figure*}[t!]
\includegraphics*[width=.9\linewidth]{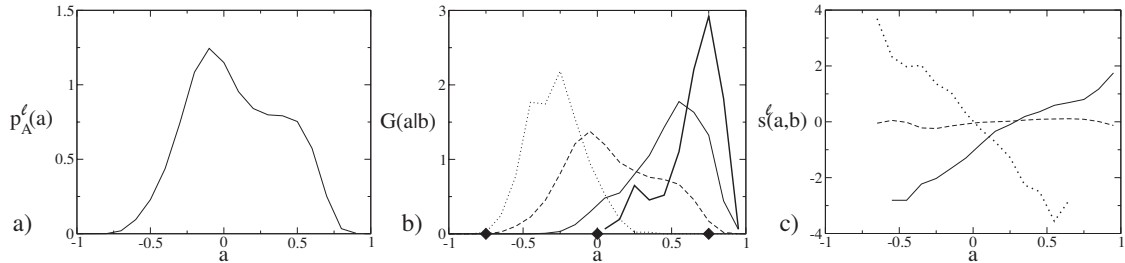}
\caption{\label{fig_evo_of_links} \small
{\bf Evolution of co-expression links between human and mouse.} 
(a) The distribution of  $p^\ell_A(a)$ of link strengths in human.  
(b) The conditional distribution of link strengths in human, $G(a|b)$, plotted against $a$
for the values $b=-0.75$ (dotted), 0 (dashed), 0.75 (full) in mouse.  The heavy solid line shows the
conditional distribution $G(a|b=0.75)$ restricted to links within expression clusters, see
text.   
(c) The empirical link scoring function $s^\ell(a,b)$ 
for $b=-0.75$ (dotted), 0 (dashed), 0.75 (full).
}
\end{figure*}

\para{Aligning human/mouse expression data.}
The co-expression networks were constructed from the expression data
of Su et al.~\cite{Su.etal:2004} obtained from 79 tissues in human, 61
tissues in mouse, and a set of biological and technical replicates of
the same size.  Similar experimental protocols were used in both
species, making the data particularly suitable for cross-species
comparison. Our networks $A$ (human) and $B$ (mouse) of size $N_A =
N_B = 2065$ contain all genes which are expressed in all samples and
show a low variance of expression levels across samples in both
species (housekeeping genes), as well as all genes having a high
expression similarity with at least one such housekeeping gene. The link
strength $a_{ii'}$ is defined as the Spearman correlation between the
expression levels of the human genes $i$ and $i'$ across all
tissues, and similarly $b_{jj'}$ in mouse. Both networks have a
broad distribution of link values; the distribution $p_A^\ell (a)$ in
human is shown in Fig.~\ref{fig_evo_of_links}(a).  To determine the
link scoring function $s^\ell(a,b)$, we look at all human gene pairs
$(i,i')$ which have homologs $(j,j')$ in mouse and compute the
distribution of link pairs $a=a_{i i'}$ and $b=b_{j j'}$. The optimal
alignment $\pi$ (along with the optimal node model parameters
$\lambda_n, \lambda_n', \zeta_0$) is then inferred by likelihood
maximization as described above; it consists of $1956$ genes.

The overall cross-species variation of expression is given 
by the root mean square difference 
$\Delta_\ell \equiv \surd {}\langle  \left(a - b \right)^2 \rangle$ (with the brackets 
$\langle \dots \rangle$ indicating the average over all aligned link pairs $a,b$), 
we find $\Delta_\ell = 0.33$. 
To separate this variation into evolutionary
changes and sampling noise, we again construct co-expression networks
from a randomly chosen subset containing half the expression
measurements from either organism and obtain $\Delta_\ell = 0.35$,
i.e., sampling contributes only a small fraction to $\Delta_\ell$. 
The alignment is also remarkably stable with respect to this change 
of the dataset: 85\% of the nodes are aligned to the same partner. 
This shows that co-expression 
networks provide a faithful representation of evolutionary changes of expression patterns. 

To trace the link evolution between the networks in more detail, we look at  
the conditional distribution 
$G(a|b)$ of correlation values 
$a_{ii'} = a$ in human given a certain correlation value $b_{\pi(i) \pi (i')} = b$ in mouse,
which is shown in Fig.~\ref{fig_evo_of_links}(b) as a function of $a$ 
for three different values of $b$.   As expected,
the variance of $G(a|b)$ is largest for weak correlations and less for strong
positive or negative correlations. The resulting link scoring function 
$s^\ell (a,b) = \log[G(a|b) / p^\ell_A (a)]$ is a linear function of $a$ with the slope 
determined by $b$, see Eq.~(\ref{Sl}) and Fig.~\ref{fig_evo_of_links}(c).

\begin{figure}[t!]
\includegraphics*[width=.95 \linewidth]{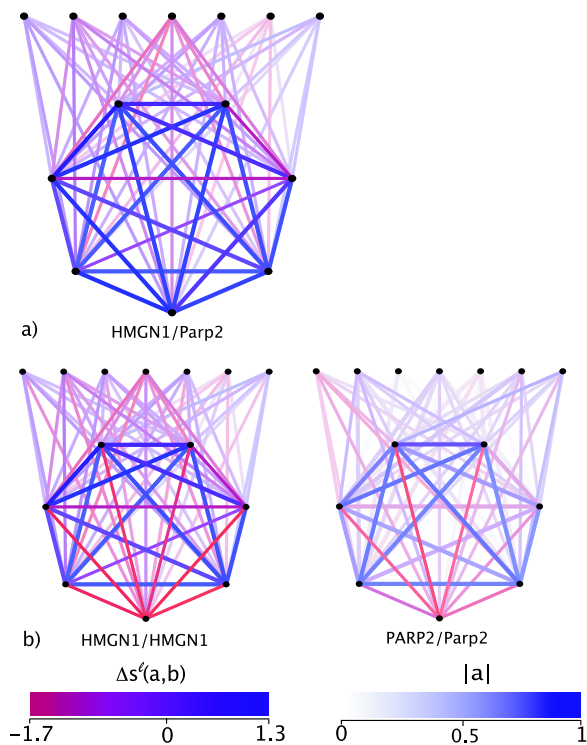}
\caption{\label{fig_netpics} \small
{\bf Evolutionary conservation of gene clusters.} 
(a) $7$ genes from a cluster of co-expressed genes 
together with $7$ random genes outside this cluster (straight line). Each node
represents a pair of aligned genes in human and mouse. The intensity and
color shading of a link encode the correlation value $a$ in human and its
relative evolutionary conservation between the two species (see color bars).
This cluster contains
the non-orthologous aligned gene pair human-HMGN1/mouse-Parp2, predicting a
non-orthologous gene displacement. 
(b) The same cluster, but with human-HMGN1 ``falsely'' aligned to its ortholog
mouse-HMGN1 (left), and human-PARP2 aligned to its ortholog mouse-Parp2 
(right, with the intensity encoding the correlation in mouse). 
This mismatch shows the poor expression similarity for this pair of genes. 
}
\end{figure}

\para{Conserved network patterns.}
Co-expression networks are not homogeneous~\cite{StuartSegalKollerKim:2003}. 
Instead, they are organized in clusters of genes which have similar expression
profiles. In the mouse network, we call a gene $j$ {\em clustered} if it has a
correlation $b_{jj'} > 0.8$  with more than 15 other genes (the average
number of links $b > 0.8$ is approximately 1 per gene). With  this definition,
there are 40 clustered genes in the network $B$ (little of the following
depends on the  exact thresholds chosen).  The thick line in
Fig.~\ref{fig_evo_of_links}(b) shows the conditional distribution
$G(a|b=0.75)$ restricted to links $b = b_{jj'}$ involving a clustered mouse
gene $j$. The root mean change of the expression correlations is 
$\Delta_\ell=0.22$. This is a reduction by a factor of two,
compared to the distribution $G(a|b=0.75)$ for all genes. 
This reduced change of expression correlation for clustered genes translates 
into a local excess link score~(\ref{local_excess_link_score}) of $\Delta S^\ell \sim 10$
per gene. This suggests that
clustered genes have  more strongly conserved expression patterns than genes
which are not part of clusters, see also ref.~\cite{StuartSegalKollerKim:2003}.  
Fig.~\ref{fig_netpics}(a) shows the link evolution between a set of
clustered genes (arranged in a circle) and a randomly chosen set of genes
outside this cluster (arranged in a straight line). The link intensity
encodes the correlation strength $a$ in human, the color its evolutionary
conservation as measured by the excess link score $\Delta s^\ell(a,b)$. 
Intra-cluster  links are, on 
average, stronger (i.e., more intense) and at the same time more conserved 
(i.e., contain more blue) than links with external genes. The genes
contained in this cluster are involved in the control of transcription and 
code for constituents of the ribosome; their  full list of 
is given as {\it Supporting Table}.

\begin{figure}[t!]
\includegraphics*[width=.95 \linewidth]{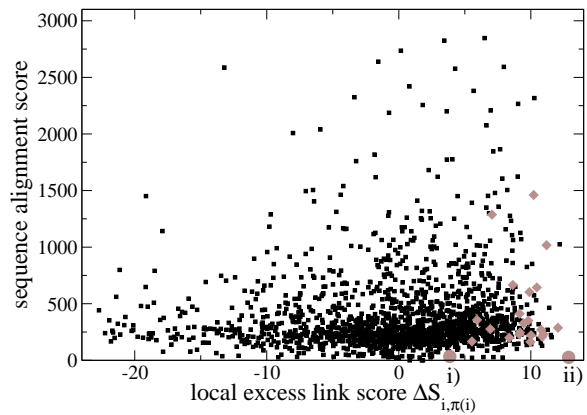}
\caption{ \small
{\bf Node versus link evolution.} \label{fig_blastvslinkscore}
For aligned pairs of genes ($i,\A(i)$), the nucleotide Blast score with 
standard parameters (vertical axis) is plotted 
against the excess link score $\Delta S^\ell_{i \pi(i)}$ (horizontal axis). 
Genes in the cluster shown in 
Fig.~\ref{fig_netpics} (gray diamonds) are distinguished by high 
link similarity, but most of them 
show no enhanced Blast score. The gene pairs (i) and (ii), aligned 
solely on the basis of the link score
(see text), are indicated by gray circles. 
}
\end{figure}




\para{Correlations between link and node similarity.} 
Figure~\ref{fig_blastvslinkscore} shows an overall correlation between
cross-species sequence similarity quantified by the score of the best 
nucleotide Blast hit~\cite{altschul.etal:1997} 
and link similarity measured by the excess link score $\Delta S^\ell$.
Gene pairs with a high sequence score also have a bias towards high link
similarity. However, the converse is not true: most of the gene pairs with 
strongly conserved expression patterns have only average sequence similarity. 
An example is the gene cluster discussed above (marked by grey diamonds
in fig.~\ref{fig_blastvslinkscore}), which has a significant excess link score  
$\Delta S^\ell \sim 10$ and a sequence score of $440$ 
per gene, which is not significantly above the network average of $394$. 

\para{Network-based gene annotations.}
Network alignment as a putative functional map differs from the homology map
of individual genes: there are genes without an (easily detected) 
homologous partner in
the other network. These genes are aligned solely on the basis of their link
score.  Although our dataset  is centered around housekeeping genes and may
be biased against such cases, the maximum-likelihood alignment contains 
significant cases of such link-based alignments, which are reported in the {\em
Supporting Table} (and marked by gray circles in
fig.~\ref{fig_blastvslinkscore}). 

\label{individual_cases}
(i) Human-OR1C1 is aligned to mouse-Olfr836 with a local  link
score $S^\ell =16.1 $ exceeding the average value $6.7$ between
orthologs.  A functional relationship between these genes is quite
plausible: Not only are both genes involved in olfactory receptor
activity \cite{GO:2000}, they also have two protein domains in common
and belong to the same gene family. However, their overall DNA
sequence identity is below $60\%$, compared to a typical value of
$85\%$ between orthologs in human and mouse. Most likely these genes are 
distant orthologs, predating the human-mouse split. 
This is an example where 
functional constraints maintain a high level of conservation at the 
network level, but not at the sequence level. 

(ii) In the case human-HMGN1/mouse-Parp2, both genes have orthologs but the 
network alignment does not match the orthology map. As shown in Fig.~\ref{fig_netpics}(a),
the human gene HMGN1 is part of a gene cluster, and the alignment to
mouse-Parp2 (with $S^\ell = 25.1$) respects the evolutionary conservation of
that cluster. The  ``false'' alignment human-HMGN1/mouse-HMGN1 respects
orthology but produces a link mismatch ($S^\ell = -12.4$) due to the poor
expression similarity of mouse-HMGN1 with the other genes of the cluster; see
Fig.~\ref{fig_netpics}(b). Human-HMGN1  is known
to be involved in chromatin modulation and to act as an
RNA-polymerase II transcription factor, in particular through
altering the accessibility of regulatory DNA. The network alignment
predicts a similar role of Parp2 in mouse, which is distinct from
its known function in the poly(ADP-ribosyl)ation of nuclear proteins. 
This prediction is consistent with a recent experimental study inhibiting the 
members of the Parp gene family in mouse. The authors conclude that
``in addition to known functions of poly(ADP-ribosyl)ation, some so
far unrecognized, non-redundant functions may also exist'',
specifically the chromatin-remodeling involved in gene expression
changes during development~\cite{Imamura:2004}.

\subsection*{Discussion}

\para{Alignment provides a quantitative measure of network divergence.}
We have developed a probabilistic alignment procedure for biological networks
based on their link and node similarity. Both components of similarity 
are important, i.e., a network alignment differs, in general, both from a mere
matching of link patterns and from a mere node homology map. To the extent
that significant sequence homology is present, it clearly introduces a
bias for the functional association of genes across organisms, and hence
for the alignment. It is this bias that makes the problem computationally 
tractable: although there is probably no formal solution by a polynomial-time
algorithm, biological network alignment allows for more efficient heuristics
than generic pattern matching. (We have discussed here an alignment 
of about 2000 genes, but ongoing studies suggest the method can be scaled 
up to genome-wide cross-species comparisons of vertebrates.)
On the other hand, the homology relations are not completely respected 
even between relatively close species: network alignment thus predicts
a deviation of functional evolution from sequence evolution for some genes.
Assessing the statistical significance of such functional swaps requires
tuning the relative weight of link and node similarity in a consistent way,
which is done here by a Bayesian inference from the datasets. 

\para{Cluster conservation and selection.}
There are important differences in the population genetics of sequences and
networks. Sequence divergence has an approximate molecular clock of
synonymous nucleotide changes, which can be described approximately by
neutral evolution. Adaptive changes can be quantified relative to neutral
evolution.  For interaction networks, the relative weights of neutral
evolution, negative and positive selection are far less clear. 
Indeed, the role of selection in the evolution of expression patterns
is currently under debate~\cite{Khaitovich.etal:2004,Yanai.etal:2005}.
Even the direction of evolution may not be as predominantly divergent as for
sequences: the selection for a given function may lead to convergent
evolution of network structures. Nevertheless, there is some regularity in
the evolution of expression patterns: genes  which are part of a strongly
correlated cluster in one species have a significantly reduced cross-species
variation of their expression profile; this conservation is quantified by a
typical excess link score $\Delta S^\ell$ of order $10$ per gene. Selection
for functionality is indeed a possible explanation. However, as the example
of Fig.~\ref{fig_netpics} shows, selection in a network can be rather complex:
conservation of a gene cluster as a whole could be attributed to purifying
selection at the level of network interactions, but this does not exclude positive 
selection leading to functional swaps at the level of network constituents. 


\para{Network-based prediction of gene function.}
Given a cross-species alignment of gene networks, we can quantify link and node
evolution. For our cross-species analysis between human  and mouse, the
correlations between these two modes are shown in 
Fig.~\ref{fig_blastvslinkscore}. Although high sequence similarity predicts 
high link conservation, most of the gene pairs with high link conservation have
only average sequence similarity. Hence, the network alignment contains 
functional information beyond the corresponding sequence alignment: it detects
evolutionary conservation which is not discernible by a comparison of overall
similarity between sequences.  Identifying genes with conserved
expression  patterns will also aid the cross-species analysis of regulatory
binding sites,  where a rapid turnover of binding sites despite the
conservation of expression  patterns has been found~\cite{Tanayetal:2005}.  

Extreme cases of mismatch between link and node evolution 
are gene pairs with significantly similar interaction patterns
but with no significant sequence similarity at all. This mismatch 
can be due to long-term sequence evolution between orthologous genes,
which randomizes their sequence similarity, while their functional roles
are more conserved. It may also arise from link dynamics leading to link
similarities between genes that are completely uncorrelated at the sequence 
level~\cite{KooninMushegianBork:1996,GalperinKoonin:1998,Morett_etal:2004}. 
In our alignment of co-expression networks, we find evidence for both 
processes. Thus, the alignment leads to functional predictions on the basis 
of network similarity alone, in cases where a functional annotation is known 
for one of the aligned genes.

\para{Acknowledgments:}
This work was supported through DFG grants SFB/TR 12, SFB 680, and 
BE 2478/2-1. We thank Terence Hwa, Daniel Barker, and Diethard Tautz 
for discussions. 


\end{document}